\documentclass[sigconf]{acmart}
\usepackage{longtable}
\usepackage{array}
\usepackage{ragged2e}
\usepackage{makecell}
\usepackage{booktabs}
\usepackage{xcolor}
\usepackage[english]{babel} 
\usepackage{blindtext}

\AtBeginDocument{%
  }

\copyrightyear{2026}
\acmYear{2026}
\setcopyright{cc}
\setcctype{by-nc-nd}
\acmConference[CHI '26]{Proceedings of the 2026 CHI Conference on Human Factors in Computing Systems}{April 13--17, 2026}{Barcelona, Spain}
\acmBooktitle{Proceedings of the 2026 CHI Conference on Human Factors in Computing Systems (CHI '26), April 13--17, 2026, Barcelona, Spain}

\acmDOI{10.1145/3772318.3791937}
\acmISBN{979-8-4007-2278-3/2026/04}

\usepackage{tablefootnote}

\begin{document}

\title[Navigating Marginalization: Toward Justice-Oriented socio-technical Design]{Navigating Marginalization: Toward Justice-Oriented socio-technical Design for Parent-Child Learning among Southeast Asian Immigrant Mothers in Taiwan}


\author{Ying-Yu Chen}
 \affiliation{%
 \institution{National Cheng Kung University}
 \city{Tainan}
 \country{Taiwan}}

\author{Yang Hong}
 \affiliation{%
   \institution{ University of Illinois Urbana-Champaign}
   \city{Urbana}
   \country{United States}
 }

 \author{Yan-Rong Chen}
 \affiliation{%
  \institution{National Yang Ming Chiao Tung University}
  \city{Hsinchu}
  \country{Taiwan}}

\author{Yi-Chieh Lee}
 \affiliation{%
   \institution{National University of Singapore}
   \city{Singapore}
   \country{Singapore}}


\begin{abstract}
This study investigates how Southeast Asian (SEA) immigrant mothers in Taiwan participate in their children’s home-based learning. Drawing on semi-structured interviews and diary studies, we explore how these mothers navigate sociocultural constraints while fostering engagement and transmitting cultural values. Despite facing diminished agency and structural marginalization, mothers engage creatively in their children's everyday learning interactions. Guided by a justice-oriented lens, we identify various harms and propose design implications for socio-technical systems that center recognition, reciprocity, and accountability in parent-child learning at the individual, familial, and societal levels. Our contribution lies in foregrounding the role of intersectional identity in parent-child learning and proposing justice-oriented design directions that support the flourishing of immigrant mothers within socio-technical systems.
\end{abstract}




\maketitle

\section{Introduction}
\textcolor{black}{Southeast Asian (SEA) immigrant mothers in Taiwan navigate a precarious intersection of migration, labor, and domestic expectations. Constituting a significant demographic shift, nearly half of the foreign women married to Taiwanese men as of 2020 originated from SEA countries, notably Vietnam, Indonesia, Thailand, the Philippines, and Cambodia \cite{zemanek_multiculturalism_2023, wang_immigration_2011}. Many of these women enter Taiwan through transnational arranged marriages, a phenomenon driven by the region's political economy of care, where women from lower socioeconomic backgrounds are recruited to fill a care and reproductive labor deficit in aging Taiwanese households \cite{liang_child-rearing_2020, Daly2002CareAsGood}. Upon "marrying in," they assume the dual burden of providing economic support for their husband’s family while serving as the primary caretakers and educators of "New Taiwanese" children \cite{Chou2010}.}

\textcolor{black}{However, this critical role conflicts with their structural position in Taiwanese society. Despite state-mandated assimilation efforts like the "New Southbound Policy" \cite{Ministry_of_Education}, these mothers often encounter an "unfriendly social environment" defined by gendered, racial, and class-based oppression \cite{Chou2010}. Within the high-stakes context of Taiwanese education, which demands "intensive parenting," they are expected to manage their children's academic success while simultaneously being stigmatized as "foreign brides" lacking the cultural capital to do so \cite{liang_child-rearing_2020, Chou2010}. This paradox creates a profound structural barrier: they are tasked with educating children within a system that systematically devalues their own linguistic and cultural knowledge.}

\textcolor{black}{Prior HCI research on parent-child learning posits the home as a collaborative space where technology can enhance bonding and literacy \cite{Druga2022, Barron2009}. Yet, these studies often assume a democratic household structure and a recognized parental authority, conditions that are frequently absent for marriage migrants living under the patriarchal surveillance of extended families. Furthermore, while migration research in HCI has examined how newcomers use technology for settlement \cite{Almohamed2016, wong-villacres_parenting_2019}, it has largely focused on short-term adaptation in Western contexts. It has engaged less with the long-term socio-technical experiences of marriage migrants in non-Western settings, whose integration is complicated not by a lack of effort, but by permanent structural othering \cite{wang_being_2010, Almohamed2016, Morales2015}.}

\textcolor{black}{To bridge the gap between the structural marginalization of SEA mothers and the potential of socio-technical design, this study investigates the dynamics of informal learning within these mother-child dyads. Through semi-structured interviews and diary studies with SEA immigrant mothers, we ask:
}

\begin{description}
   \item[\normalfont RQ1:] How do SEA immigrant mothers in Taiwan engage in learning activities with their children at home? What socio-cultural factors enable or constrain their engagement in this regard?
   \item[\normalfont RQ2:] What challenges do SEA immigrant mothers face, and what resources and technologies do they \textcolor{black}{employ} when engaging in their children’s learning?
   \item[\normalfont RQ3:] How might socio-technical systems be designed to support SEA immigrant mothers' engagement in their children's learning at home?
\end{description}

\textcolor{black}{Our findings reveal that SEA mothers engage in creative, often invisible forms of cultural transmission to counter their marginalization, such as the oral sharing of heritage folktales and the maintenance of mother-tongue languages within the household. However, they struggle with feelings of inadequacy and relational silencing where their children and in-laws dismiss their competence. While they actively use technology to support schoolwork, existing tools often reinforce the dominant curriculum that excludes them.}

\textcolor{black}{We respond to these findings by proposing an \textit{Intersectional Justice-Oriented Design} framework. We argue that addressing the disconnect between migrant experiences and technology requires moving beyond deficit-based design toward structural responsiveness. Our contribution is threefold: First,  we synthesize perspectives from migration studies and education technology to frame the home not as a neutral learning space, but as a political site where "care" is shaped by the political economy of marriage migration. Second, we articulate the specific harms SEA mothers face—specifically the erasure of their cultural competence and patriarchal restrictions on their agency—and how current technologies fail to mitigate them. Third, we advance a justice-oriented framework that maps design interventions to structural realities. We propose designing for \textit{competence visibility} at the individual level, \textit{subversive reciprocity} at the familial level, and \textit{structural accountability} at the societal level. The \textit{Intersectional Justice-Oriented Design} framework actively addresses the systemic conditions of marginalization, aiming for the flourishing of immigrant mothers within socio-technical systems \cite{kirabo2021inclusive, crenshaw2017intersectionality, Schlesinger2017}. It serves as a scaffold to support researchers and practitioners in navigating these structural constraints to design for intersectional justice.}

\section{Related Work}
In this section, we situated SEA immigrant mothers’ parenting experiences in Taiwanese contexts and traced shifts in migration research in HCI. We also examine parent-child learning in underrepresented families and how marginalization operates across multiple levels.

\subsection{\textcolor{black}{The Structural Context of Motherhood: Immigrant Marginalization and Educational Pressure}}

\textcolor{black}{To understand the lived experiences of SEA immigrant mothers, we must first situate them within the political economy of care—a theoretical framework that analyzes how reproductive labor (cooking, cleaning, child-rearing) is distributed, valued, and commodified across global borders \cite{Daly2002CareAsGood, Razavi2007PoliticalSocialEconomy}. In the context of East Asia’s "care deficit," caused by aging populations and increasing female workforce participation, care work is frequently outsourced to migrant women through transnational marriage \cite{Parrenas2001Servants, WangChang2002CommodificationMarriages}. Scholars argue that this structure renders immigrant wives as "commodities of care," valued primarily for their biological and domestic utility rather than their cultural or intellectual citizenship \cite{Lan2003GlobalCinderellas}. Consequently, their presence in the Taiwanese household is shaped by a "global care chain" that extracts emotional and physical labor while structurally marginalizing their social status \cite{Hochschild2015GlobalCareChains}.}

\textcolor{black}{These structural vulnerabilities are compounded by Taiwan’s education system, which is deeply rooted in Confucian heritage culture, where academic achievement is viewed as a collective family project and a primary indicator of moral character, rather than solely an individual pursuit \cite{Wanless2015Taiwanese, Stankov2010Unforgiving}. Within this high-stakes, credentialist environment, parents are expected to practice "intensive parenting" or "concerted cultivation," actively managing their children's schedules and academic resources to ensure social mobility \cite{KAO2021102535}. This dynamic institutionalizes the home as an extension of the school, most visibly through the home-school communication book. In this mandatory daily log, teachers and parents exchange notes on homework, grades, and behavior \cite{Chen2019Understanding}. This system structuralizes parental involvement, requiring parents to act not just as caregivers but as "assistant teachers" who monitor and enforce academic discipline daily \cite{Hung2007Family}.}

\textcolor{black}{Crucially, these responsibilities are disproportionately gendered, constructing the "good mother" in Taiwan as an "education manager" whose own social worth is tethered to her child's success \cite{KAO2021102535}. While recent reforms, such as the 12-Year Basic Education curriculum, have shifted rhetoric from rote memorization to "holistic development" (\textit{su-yang}) and critical thinking \cite{MOE2020aSDGManual}, scholars argue this has paradoxically increased the burden on parental labor \cite{Kao2001Childrens, TaipeiTimes2025ParentsFallingBehind}. Mothers are now expected to cultivate not only test scores but also diverse competencies and moral virtues (\textit{pin-ge}), a role that demands significant cultural capital and linguistic proficiency in Mandarin \cite{ShenYuan1998MoralValues, lin2016facebook}. For immigrant mothers, this intensified, surveillance-heavy model of domestic learning often contrasts sharply with the educational norms of their home countries, creating a structural barrier where their alternative forms of care are rendered invisible or inadequate by the dominant school system \cite{Chen2025ParentInvolvementTaiwan, zemanek_multiculturalism_2023, Chou2010}.}

\textcolor{black}{Existing research confirms that these systemic pressures profoundly shape their parenting experiences. Language and cultural gaps often hinder communication with teachers and healthcare professionals, as the expectation of "intensive parenting" clashes with their linguistic realities \cite{lan_reproductive_2019, Lan2003GlobalCinderellas}. Chou examined how immigrant status affects psychological well-being and children’s development \cite{Chou2010}, while Tsai and Lee noted that limited interpersonal and linguistic competence constrains SEA mothers’ interactions with public institutions \cite{Tsai2016}. Liang et al. further observed that unfamiliarity with Taiwan’s education system and limited formal schooling restrict mothers’ capacity to assist children academically \cite{liang_child-rearing_2020}. Other studies echo these findings, identifying difficulties in navigating curricula and accessing educational resources or social support networks \cite{wang_being_2010, hsia2000transnational, bornstein2004sitting}.}

\subsection{Migration Research in HCI}
HCI scholarship has long examined how immigrants navigate sociocultural and institutional barriers, often through ICTs. Much of this work focuses on newcomers in Western contexts, where technology is framed as a mediator that supports settlement through access to resources, communication, and services (e.g., \cite{Almohamed2016, Hsiao2018, Rohanifar2022, Bishop2015, wong-villacres_parenting_2019}). \textcolor{black}{Studies highlight issues such as language, financial exclusion, and healthcare access, showing how immigrants adapt technologies or adopt new systems to bridge gaps \cite{Truong2024, Sabie2022, Brown2014, Bletscher2020Communication}.} For instance, Hsiao and Dillahunt \cite{Hsiao2018} found that immigrant parents relied on peer forums to identify trusted resources. \textcolor{black}{Berg (2022) explores how refugee women in Germany used mobile phones and digital media to overcome information precarity. Technologies like mobile phones functioned as essential lifelines for refugee families to access information, social networks, and resources \cite{Shah2019Family, Kuneva2023Fostering, CorreaVelez2020Social}.}

Yet, most of this work centers on newcomers’ practical needs, with less attention to the social and emotional experiences of long-settled migrants whose interactions with technology remain shaped by structural exclusion \textcolor{black}{whose interactions with technology remain shaped by structural exclusion \cite{Tachtler2021, Sabie2022, Hsiao2023How}. Besides, there is relatively little research on immigrant Southeast Asian women in Global North Asian countries. Ethnographic studies of migrant domestic women in countries such as Singapore, for example Indonesian domestic workers, show that ICTs can play a critical role in maintaining social ties, reducing isolation, and negotiating mobility and social integration \cite{Platt2016Renegotiating, ThomasLim2011MaidsMobile, Platt2013FinancingMigration}.}

Recent work has expanded to the long-term lives of immigrants, emphasizing cultural identity, intergenerational dynamics, and emotional well-being \cite{Sabie2022, Truong2024, Zhao2023}. However, families are often treated as cohesive units, obscuring individual roles and constraints within households \cite{Glick2010, Vancea2013}. Migrant women are frequently positioned as “domestic actors,” limiting recognition of their agency and reinforcing marginalization \cite{Sabie2022, BanulescuBogdan2020}. Responding to calls for more intersectional approaches \cite{Sabie2022}, the study examines Southeast Asian immigrant mothers in Taiwan, many of whom have lived in the country for over a decade but continue to face linguistic, cultural, and economic barriers \cite{Liang2020, Chou2010}. Extending perspectives on “non-use” as socially and structurally shaped \cite{baumer_why_2015, sultana_design_2018}, this study investigates how these mothers engage in their children’s learning at home, both with and without technology, and how their strategies reflect and resist everyday marginalization.

\subsection{Parent-Child Learning in Underrepresented Families}

Parental engagement in children's learning has been associated with a wide range of developmental benefits, including stronger family connectedness and emotional bonding \cite{Padilla_Walker2012}, improved cognitive and social skills \cite{barron2009parents}, increased self-efficacy \cite{landry2006influence}, and enhanced academic performance \cite{Jeynes2005}. Collaborative learning in families—especially in creative computing and tech-mediated environments—often involves shifting roles, where parents act as facilitators and children as instructors \cite{Takeuchi2011, Martinez2022, Banerjee2018}. Such practices enable shared problem-solving and identity formation through joint exploration \cite{Banerjee2018, Yu2023, Yu2021}.

\textcolor{black}{However, in underrepresented and immigrant families, structural barriers like limited language proficiency\cite{Liaqat2021Intersectional}, low education levels, and unfamiliarity with school systems constrain parents’ involvement \cite{Golinkoff2019, Liaqat2021, Behnke2008}. }Educational tools often assume English fluency and cultural familiarity, sidelining parents from active participation \cite{Vikki_kids_2015, Banerjee2018}. Even well-intentioned family-oriented coding initiatives can pose accessibility issues when they rely on English-dominant systems, like Scratch \cite{Resnick2009} in the Family Creative Learning program \cite{peppler_makeology_2016}.  In these contexts, immigrant mothers may defer learning support to schools or siblings and feel excluded from their children’s educational experiences \cite{Wong_Villacres_Work_2019}.

To address these challenges, \textcolor{black}{HCI scholars have explored technologies that support not just learning content but also negotiation, co-creation, and affective dynamics in family learning \cite{Takeuchi2011, Yu2023, Wyche2012This}.} Systems like Captivate \cite{Kwon2022} and BlockStudio \cite{Banerjee2018} support contextual language learning and co-creation interaction. Meanwhile, researchers have examined the different roles parents take during children’s learning activities, such as teacher, project collaborator, or scaffolder, and how these roles shift dynamically depending on media and contexts \cite{Yu2023, Druga2022, barron2009parents}. Yet, little is known about how such dynamics play a role in non-Western immigrant households. Our study explores these dynamics in the context of SEA immigrant mothers in Taiwan, seeking to understand how they engage in their children's learning activities. 

\subsection{Marginalized Population and Social Justice in HCI}
Social justice has increasingly become a critical lens in HCI for examining how power, privilege, and access are distributed in socio-technical systems \cite{jost_social_2010, nussbaum__beyond_2004}. Early definitions centered on equitable distribution of resources within Western liberal traditions \cite{rawls_theory_1971}, but scholars have critiqued these frameworks for overlooking the structural oppression of non-white and non-Western communities \cite{mills_rawls_2009, ogbonnaya-ogburu_critical_2020, erete_i_2021}. This exclusion has prompted critiques of mainstream social justice theories, arguing for a more inclusive approach that recognizes the structural oppression experienced by non-white communities. In response, HCI researchers have proposed justice as an evolving, situated framework for addressing lived experiences of marginalization and oppression through technology design \cite{bardzell_feminist_2010, fox_exploring_2016, desportes_examining_2021, corbett_engaging_2019, dombrowski_social_2016, bellini_there_2022}. Others have extended these frameworks by examining how harms emerge and how justice/injustice is conceptualized in the literature \cite{corbett_engaging_2019, chordia_social_2024}, arguing for reflexive practices that respond to injustice across individual, community, and structural levels. Dombrowski et al. adapted Lötter’s six dimensions (transformation, recognition, enablement, reciprocity, distribution, and accountability) to guide justice-oriented design \cite{dombrowski_social_2016}. This perspective pushes designers to move beyond merely acknowledging inequality and to actively redesign systems that more equitably serve underrepresented and marginalized groups in non-Western contexts \cite{dombrowski_social_2016, fox_social_2017}. 

This convergence and employment of social justice frameworks has led HCI researchers to more critically analyze how technology may perpetuate or challenge inequities among marginalized populations \cite{ma_advancing_2024, munoz-alcantara_justice-led_2024, xiao_sensemaking_2022}. For instance, data-driven systems in healthcare have been found to undermine the autonomy of marginalized care workers, reinforcing stereotypes and diminishing agency, which indicates a need for technologies that prioritize dignity and respect \cite{sun_care_2023, poon_computer-mediated_2021}. In another example, Strohmayer et al., adopting an intersectional restorative justice lens, collaborated with a Canadian sex worker rights organization to explore technology's role in reporting violence, thereby enhancing safety for this stigmatized community \cite{strohmayer_technologies_2019}.

Framed by postcolonial computing, Irani et al. argue that justice in global contexts must grapple with power, authority, and participation \cite{irani_postcolonial_2010}. The research in the non-Western context investigates the experiences of SEA immigrant mothers and uncovers various challenges that hinder their engagement in children’s learning. It reflects on their current use of technologies in the learning process, considering design implications beyond addressing immediate functional needs, and empowers the marginalized users to navigate the structural constraints for self-actualization.

\section{Method}
Data was collected from fourteen SEA immigrant mothers in Taiwan through semi-structured interviews and a two-week diary study in spring 2022. With participants’ consent, all interviews were audio-recorded and transcribed verbatim, each lasting between 1 and 2 hours. During the diary phase, participants contributed multimedia materials including 78 videos, 44 audio clips, and 385 photographs with associated narratives, which allowed the researchers to gain a comprehensive understanding of participants’ daily practices in real-life settings.

\subsection{Study Procedure}
Our research process consisted of three phases: (1) an initial semi-structured interview, (2) a two-week diary, and (3) a follow-up semi-structured interview. In the first phase, we interviewed participants to understand their backgrounds, parenting values and how they make sense of home-based learning activities with their children. We asked about the challenges mothers encountered in participating in their children’s learning and family life, as well as the technological and non-technological strategies they used to address them. During the second phase, participants kept a two-week diary via Messenger, a social media platform familiar to them. In this process, participants were reminded to spend around 10 minutes each day recording the moments or events that they engaged in learning activities with their children, and they were encouraged to supplement text with photos, videos, and audio clips. The follow-up interview in the third phase built on insights from the initial interview and diaries, allowing us to probe further details and ensure accurate interpretation of the data. Thirteen participants completed all three phases, while one (P1) participated only in the first interview due to her scheduling conflicts. The median number of diary entries was 13 days, with an average of 12.5 days per participant. Each participant recorded 30 photos, 6 videos, and 4 audio clips on average.

Before the first interview, we explained the study procedures and obtained informed consent from participants. We provided each participant with a bilingual consent form in Mandarin and their native language to support comprehension. To maintain ongoing consent, we regularly reminded participants of their right to withdraw at any time in all three phases. All procedures and protocols received approval from our university’s Institutional Review Board.

\begin{figure*}
  \centering
  \includegraphics[width=\linewidth]{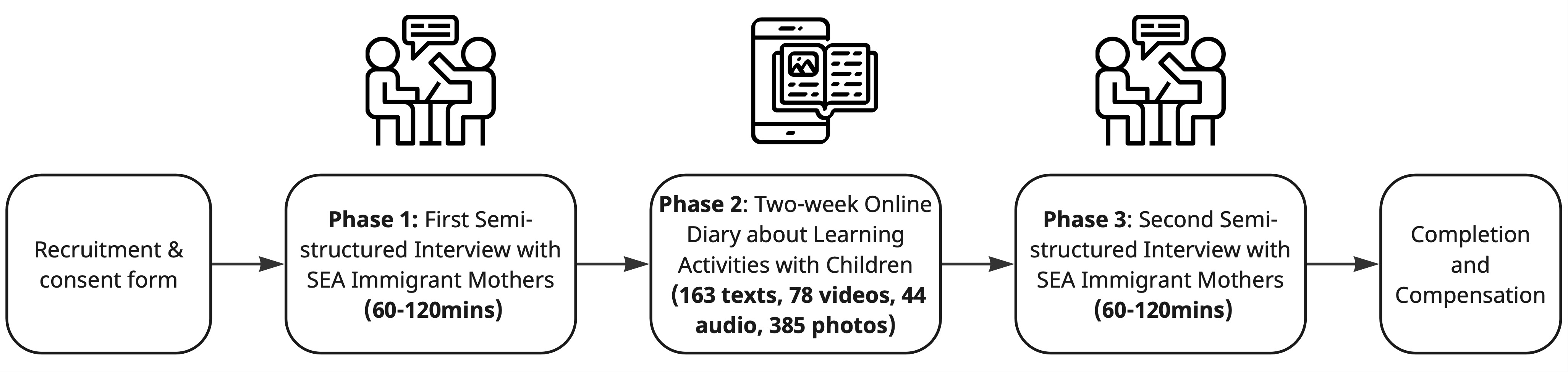}
  \caption{An overview of study procedure: consisted of two semi-structured interviews and an online diary activity}
  \Description{research procedures including three phases}
\end{figure*}

\subsection{Participants}
\textcolor{black}{We recruited participants through a combination of outreach to local organizations supporting SEA migrants in Taiwan, alongside targeted postings in relevant community social media groups (e.g., Facebook and LINE forums for immigrants) and in-person distribution of printed flyers at local immigrant family service centers.} Eligibility criteria required participants to be from Southeast Asia and mothers of at least one child aged 7 to 12 (the elementary school age range in Taiwan). This focus on mothers of younger children reflected the importance of family bonds and home-based parent-child learning at this stage. To ensure effective communication and minimize language comprehension issues, participants also needed basic proficiency in Mandarin. 

Of the fourteen participants enrolled, ten were originally from Vietnam, \textcolor{black}{three from Indonesia}, and one from Cambodia, with residence durations in Taiwan ranging from eight to twenty-one years. Thirteen participants held dual roles as both paid employees and primary caregivers for their children, while one was a housewife. Thirteen participants were married to Taiwanese men, and one to a Vietnamese man. Most participants had completed high school in their home countries (10 participants), with three having attended university and one completing junior high school. Participants primarily communicated with their children in Mandarin. Among the eight participants who disclosed household income, six reported annual earnings below NT\$500,000 (approximately US\$15,675), and two between NT\$500,000 and NT\$1,000,000. For context, the median household income in Taiwan in 2022 was NT\$940,000 (approximately US\$29,500) \cite{national_statistical_bureau_directorate}. \textcolor{black}{As shown in Table 1, all participants had lived in Taiwan for 8-21 years by 2022, indicating they were long-settled immigrants.}

\begin{table*}[h]
  \caption{Participant Demographics}
  \label{tab:participants}
  \centering  
  
  \begin{tabular}{cccc>{\centering\arraybackslash}p{1.8cm} >{\centering\arraybackslash}p{2.5cm}}
    \toprule
    Participants & Nationality & Ethnicity & Age & Number of Children & Duration of Residence in Taiwan (years)\textsuperscript{1} \\
    \midrule
    P1  & Indonesia & Chinese    & 42 & 2 & 20 \\
    P2  & Cambodia  & Khmer      & 36 & 3 & 8  \\
    P3  & Indonesia & Chinese    & 41 & 2 & 16 \\
    P4  & Indonesia & Chinese    & 39 & 2 & 21 \\
    P5  & Vietnam   & Hui People & 41 & 3 & 20 \\
    P6  & Vietnam   & Chinese    & 35 & 2 & 8  \\
    P7  & Vietnam   & Hui People & 36 & 4 & 16 \\
    P8  & Vietnam   & Hui People & 33 & 1 & 10 \\
    P9  & Vietnam   & Hui People & 42 & 2 & 10 \\
    P10 & Vietnam   & Hui People & 42 & 2 & 17 \\
    P11 & Vietnam   & Hui People & 38 & 2 & 11 \\
    P12 & Vietnam   & Hui People & 39 & 3 & 18 \\
    P13 & Vietnam   & Hui People & 42 & 3 & 20 \\
    P14 & Vietnam   & Hui People & 40 & 2 & 19 \\
    \bottomrule
  \end{tabular}

  \par  
  \vspace{1ex} 
  
  \begin{minipage}{0.8\linewidth} 
    \footnotesize
    \textsuperscript{1} “Duration of Residence in Taiwan” reflects participants’ self-reported years of residence during interviews. Data was collected in 2022.
  \end{minipage}
\end{table*}

\subsection{Data Analysis}
We conducted a thematic analysis of interview transcripts and diary data, guided by Braun and Clarke’s reflexive Thematic Analysis (TA) method, which employs a bottom-up approach that allows themes to emerge organically through coding, without relying on a predetermined framework \cite{Braun2006}. Interviews were conducted, recorded, and transcribed in Mandarin by native-speaking research team members. Throughout the analysis, we used memos to capture deeper insights and support a thorough exploration of the raw data\cite{braun_reflecting_2019}. Our data set included 27 audio files from semi-structured interviews with our participants. Interviews were conducted, audio-recorded, and transcribed in Mandarin by native-speaking research team members. After compiling the data, we developed a codebook through an iterative process, categorizing interconnected themes to clarify and organize the relationships among them. Weekly team meetings provided a platform for discussing memo interpretations and rigorously examining, refining, and revising themes. This iterative, collaborative approach helped ensure a cohesive understanding and consistent interpretation of the dataset’s nuanced components. We include a summary of the final codebook with 23 codenames, which can be found in Appendix B.

\subsection{Positionality}
All members of our research group, except the second author who is from mainland China, were born and raised in Taiwan. We are experienced in conducting HCI research and share a strong commitment to designing technologies with marginalized communities, which motivated this study. Three authors identify as female and one as male. Three authors also received HCI training in the United States. We acknowledge that our backgrounds shape both our research questions and our interpretation of findings, particularly regarding power dynamics \cite{harrington_deconstructing_2019}. This study represents a first step in a longer-term inquiry into technology use and design among these families.

\section{Findings}
\subsection{Understanding SEA Immigrant Mothers’ Engagements in Children's Learning}

Participants described a wide range of learning-related activities they engaged in with their children at home, extending far beyond formal schoolwork. These included sharing cultural traditions, assigning household chores, participating in physical and recreational play, and fostering moral development. While participants did see academic support as a key aspect of parenting, their broader understanding of “learning” also encompassed the transmission of values, life skills, and cultural identity.

When helping with schoolwork, most mothers actively used digital tools, such as translation apps, online searches, and instant messages, to compensate for language and cultural gaps. However, these tools often failed to enable meaningful, instructional support. As a result, these mothers’ roles in school-related learning tended to be supervisory from the sidelines, rather than directly instructional. On the other hand, in non-academic learning activities (e.g., cooking traditional dishes, playing board games), participants felt more confident and competent. These practices allowed for more reciprocal and culturally grounded interactions with their children, yet were rarely mediated by technologies.

\subsubsection{Accompanying children to complete schoolwork}
Accompanying children during homework time was a common practice among all participants. However, due to differences in language, culture, and educational systems between Taiwan and their countries of origin, many mothers struggled to provide direct academic support. To bridge these gaps, they frequently relied on accessible digital tools, like smartphones, tablets, and computers, to search for information, translate school materials, and communicate with others. Yet despite their proactive use of technology, these tools often proved inadequate in addressing the deeper linguistic and cultural challenges they faced.

A major barrier was difficulty with written Chinese. While most mothers could communicate orally in Mandarin, many lacked confidence in reading or writing Chinese, even at an elementary level. Participants described using dictionary or translation apps to decipher unfamiliar words and idioms (P1, P3, P6, P10, P13), yet these methods were often slow and imprecise. P4, for example, used handwriting input to trace Chinese characters, but found the process inefficient and frustrating. Some mothers, like P5, also noted challenges in subjects like math, where instructional styles and terminologies in Taiwan differed from those in their home countries:

\begin{quote}
    “The way math is taught in Taiwan is different from how I learned it in my childhood; even the mathematical formulas are shown in a different way! … I can get the answer to the question, but I can’t just give the answer to my kid without a clear explanation. They won’t learn anything that way!”
\end{quote}

Participants used translation apps with text or photo input to help with unfamiliar words and phrases, but the translators sometimes did not work well. Participants noted inaccuracies, especially with idiomatic expressions or specialized vocabulary (P1–P5, P8, P10). As P10 reflected:

\begin{quote}
\sloppy
    “When something is translated into Vietnamese, the meaning often changes slightly, especially for idioms… Even if I get the Chinese explanation, I still need to know how to pronounce it to ensure that I understand the meaning.”
\end{quote}

These limitations often led mothers to seek help from family members. P8, for instance, used Google Translate to interpret her child’s home-school communication book but still needed clarification from her husband, as such communication practices were unfamiliar from her own schooling in Vietnam.

Due to these difficulties, many mothers reported that their engagement with schoolwork was primarily supervisory rather than instructional (P2–P9, P12). Still, they viewed this form of support as essential. Eight participants expressed that being physically present helped their children stay focused and disciplined during homework time. As P12 shared:

\begin{quote}
    “I need to stay close and watch them to ensure they concentrate on homework.”
\end{quote}

When academic issues arose, some mothers reached out to more knowledgeable family members (usually their husbands or older siblings) via instant messaging apps like LINE, Messenger, or WhatsApp (P2, P4, P6, P10, P12). Others postponed resolving the issue until someone else could help, or reminded their children to consult tutors after school (P2, P8, P10). For example, P2 used LINE to send photos of completed homework to her husband for review and correction.

In sum, although technology played a prominent role in mothers’ efforts to support schoolwork, it did not empower them to become instructors of children's homework. Most participants' engagement with children's schoolwork remained supervisory and frequently required support from other family members.

\subsubsection{Engaging children in cultural learning.}
Cultural learning, drawn from our participants' personal experience and heritage, emerged as a key activity where mothers actively engaged with their children. Across diaries and interviews, ten participants mentioned cultural learning 24 times, underscoring its importance in their daily parenting practices.

Cooking was the most frequently cited activity for passing on cultural knowledge, mentioned twelve times by six participants (P2, P3, P5, P8, P9, P11). The kitchen served as a practical and educational space, where mothers introduced traditional dishes while encouraging their children to participate hands-on. Unlike their experiences with schoolwork, participants did not require digital assistance in these activities, and they felt competent and confident sharing their own knowledge (P3, P5, P11). P11 described teaching her son to make \textit{bánh bao} (Vietnamese steamed pork buns) together as a joyful experience with the sense of accomplishment:

\begin{quote}
    “I let the kid shape the bánh bao with his hands, giving them the freedom to be creative and learn about our traditional food. We all felt very happy and accomplished to see what we made. The process is so relaxing!”
\end{quote}

Language learning was another important part of children's cultural engagement. Six participants described supporting their children in learning their mother tongue by purchasing textbooks and working through the content together (P3, P6, P7, P9, P12, P14). For example, P9 and P12 frequently shared diary entries about guiding their children’s Vietnamese studies, answering questions, and offering clarification when needed. P12 also made a conscious effort to speak Vietnamese with their children in daily life.
Beyond planned activities like cooking or studying, participants also wove cultural education into everyday routines (P3, P8, P12, P14). P14 described how she integrated Vietnamese traditions into her children's lives in informal, personal ways. After her children finished homework, she introduced them to Vietnamese board games and took time to explain cultural practices from home. For example, during Ching Ming Festival (Tomb Sweeping Day), she discussed the Vietnamese approach to honoring ancestors with her child:

\begin{quote}
    “I want my child to understand that it’s not only Taiwan that has these customs. Vietnam also celebrates Ching Ming Festival. I hope he can learn how we honor our ancestors back home. Many traditions between the two cultures are quite similar.”
\end{quote}

In these examples, mothers acted as educators who passed on their own wisdom. Participants shared that these learning activities allowed for mutual enjoyment and bonding between mothers and children (P9, P11, P12, P14).

\subsubsection{Doing chores and physical activities with children.}
In addition to schoolwork and cultural education, participants also regarded everyday activities, such as household chores, games, and physical exercises, as meaningful opportunities for learning. Participants played active roles by initiating the activities, assigning tasks, monitoring progress, or directly participating alongside their children.

Household chores emerged as the most frequently documented form of non-school learning, with 20 diary entries referencing this practice (P2, P3, P5, P7–P9, P11). Mothers viewed chore assignments as chances to cultivate life skills such as independence, time management, and responsibility. Commonly delegated tasks included washing dishes, folding laundry, vacuuming, cooking, and hanging clothes. Depending on the child’s age and ability, mothers either observed, checked the outcomes, or worked alongside their children to complete the tasks. Beyond practical skill development, completing chores together also served as moments of parent-child bonding. Several participants (P2, P3, P5, P7, P8) shared how they \textcolor{black}{spend} time together through doing chores collaboratively. As P5 described:

\begin{quote}
    “My older daughter was cleaning her computer keyboard, the younger one was cleaning the screen, while I was taking care of the seating area… We worked together as a team to complete the deep cleaning.”
\end{quote}

Besides, Mothers also saw play and physical activity as learning opportunities for new skills and all-round development. Ten participants described 13 types of outdoor activities in their diaries (P3, P4, P6–P12, P13), including visiting parks, picking oranges, beach exploration, hiking, and family camping. They mentioned that these excursions offered their children the chance to interact with nature and learn through observation and experience. Indoor recreational activities were also frequently mentioned (11 diary entries), like crafting or playing with toys. Mothers were often present, either watching their children play, setting up the environment, or joining the activity themselves. For instance, P10 described a scene in which her children were working on a dollhouse while watching TV with their father, with her nearby helping to organize the pieces.

\subsubsection{Cultivating children's virtues and good moral qualities}
Six participants shared how they actively engaged in teaching virtues and values to their children, using everyday incidents as teachable moments (P3, P6, P7, P9, P11, P12). The values they emphasized included honesty, helpfulness, sharing, politeness, patience, frugality, and safety awareness. Honesty, in particular, was described as a priority across participants’ accounts.

For instance, P3 shared a moment of pride when her child voluntarily corrected a grading error, changing a score from 97 to 91. She expressed admiration for her child's integrity, valuing honesty over academic achievement. In contrast, P6 described her disappointment upon discovering her child had lied. In response, she recounted the fable \textit{the Boy Who Cried Wolf} to emphasize the consequences of dishonesty and even recorded the issue in the home-school communication book to enlist the teacher’s support.

Participants also described using everyday situations to model and reinforce behaviors such as sharing and helping others. P6 praised her daughter for bringing extra juice from after-school care to share with her sibling. Similarly, P12 said she routinely checked whether her child had helped their grandmother with chores after visits.

Safety awareness was another recurring theme, mentioned in six diary entries. After hearing about conflicts involving classmates, P7 and P9 spoke with their children about handling such situations with care. They advised their children to remain calm if bullied and to seek help from teachers instead of reacting impulsively.

\begin{figure*}
  \centering
  \includegraphics[width=0.6\linewidth]{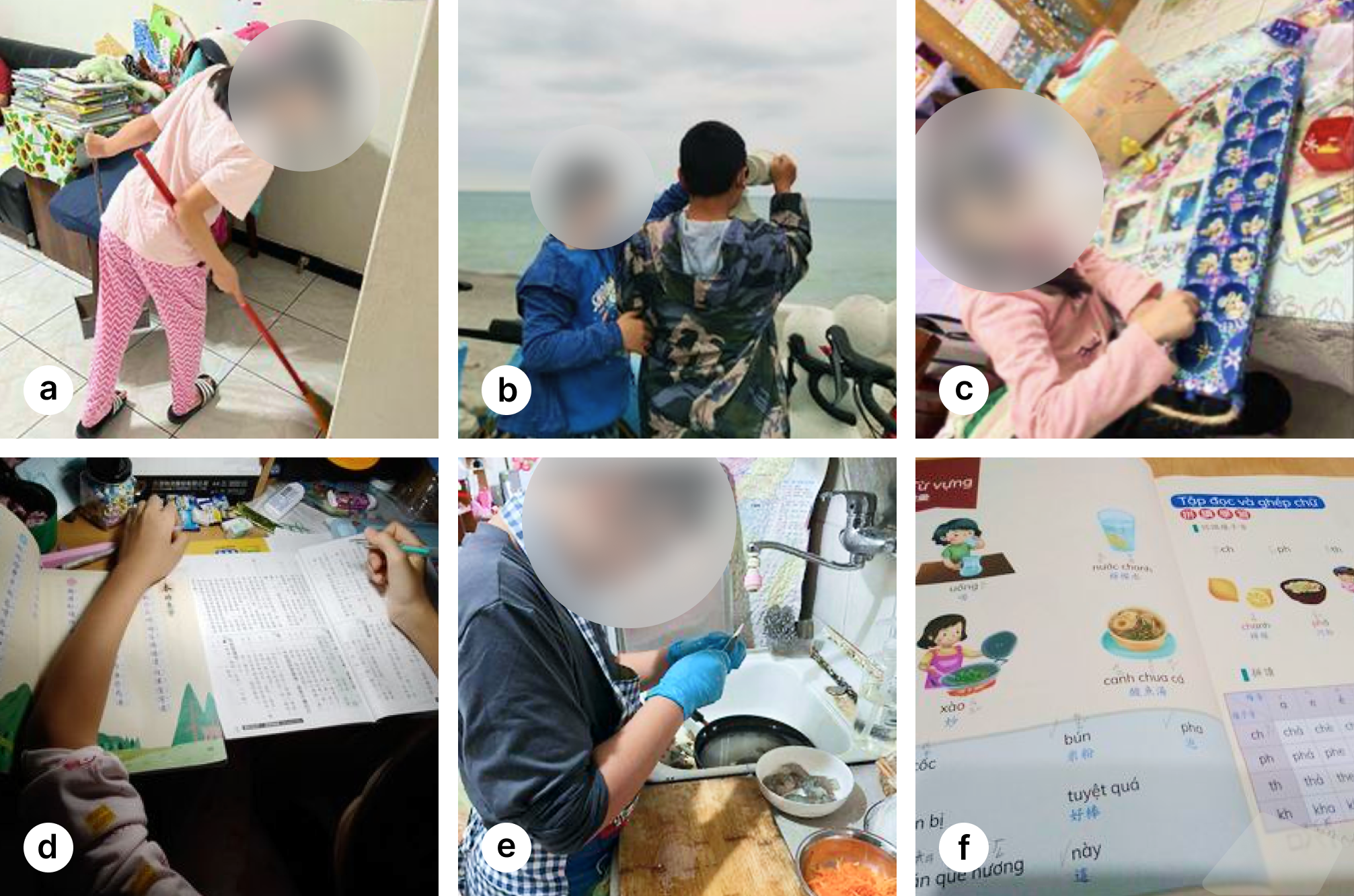}
  \caption{Photos taken by participants during the two-week diary period, illustrating (a) teaching children sweep the floor; (b) taking children to the beach to learn more about the sea; (c) playing the traditional Vietnamese game \textit{Ôănquan} with children; (d) watching children finishing schoolwork; (e) cooking with children; (f) teaching children Vietnamese with textbook at home.}
  \Description{a set of photos taken by participants about their parent-child learning activities at home.}
\end{figure*}
~
\subsection{Socio-Cultural Barriers and Emotional Struggles Faced by SEA Immigrant Mothers}
Despite participants’ efforts to stay involved in their children’s learning with or without digital technologies, they expressed different levels of frustration and emotional discouragement in this regard. Their attempts to support children’s learning at home were often constrained by deeper socio-cultural challenges that influenced their confidence, self-identity and recognition in everyday lives.

\subsubsection{Feelings of inadequacy}
Nine of our participants (P2-P5, P8-P12) said they did not feel confident about helping their children with their homework. A recurring theme was that their children often bypassed them when seeking academic help, preferring instead to consult their fathers or other family members. P1, P2 and P5 mentioned that they felt disheartened when their children almost always turned to their father for such help. P2 said she sometimes asked her children: “Why don’t you ask me? I can help you, too.” In some cases, the children not only ignored their mothers’ willingness to help with homework, but criticized them for poor Chinese pronunciation (P5), “wrong” English accents (P1), or steps solving mathematical problems that differed from those taught at school (P4). Several participants said they were saddened by their children’s attitude, and felt looked down upon by them. P2 recounted a painful experience in which her child mocked her for not understanding the home-school communication book,

\begin{quote}
    “My oldest daughter said, ‘Why don’t you know anything I ask you?’ I’m so sad and told her, ‘Mommy didn’t go to school in Taiwan. That’s why sometimes I don’t understand – because I didn’t study here! I studied in Cambodia, and our schools don’t use the same way to communicate with parents.” 
\end{quote}

In short, our participants actively sought to engage in their children’s academic activities, but their children—whether influenced by prevailing societal norms, their own perceptions, or a combination of the two—subtly dismissed and undermined these efforts. This underscores the nuanced complexities that these mothers encounter when striving for learning engagement within their familial contexts.

\subsubsection{Feeling unsupported and marginalized in the household}
The opposition our participants faced within their homes was not just from their children. Notably, their husbands and husbands’ relatives tended to disapprove of their wishes to speak their native language to their children, to train them in daily routine tasks, and to discipline them. Seven participants (P2-P5, P9-P11) stated that their mothers-in-law forbade them from speaking to their children in their native language, on the grounds that children should not learn too many languages. And when six participants (P2-P5, P9, P11) wanted to involve their children in domestic tasks, their mothers-in-law or fathers-in-law stopped them, arguing that children should not do housework.

P5 mentioned that when she asked her children to help with chores or learn their native language, her in-laws often interrupted her.

\begin{quote}
    “My in-laws believe that immigrant women marrying into Taiwanese families are here to produce offspring and serve the other household members, and I know they are not alone in thinking this way,” she said. “They also believe that daughters-in-law from Southeast Asia will take the money and send it back to their own country, and thus that we should not be treated nicely.” 
\end{quote}

Similarly, P9 told us that when she disagreed with her husband or in-laws over child-discipline matters, she had \textcolor{black}{learned} to keep her mouth shut because she was “just a foreign spouse whose opinion doesn’t matter.” All our participants shared the belief that they had little to no say in how their children were educated within their families.

\subsubsection{Feeling undervalued and unrecognized in society}
When asked to reflect on their experiences in Taiwan regarding identity, social integration, and local perceptions, several participants recounted moments of overt and subtle discrimination. Both P2 and P4 shared that strangers had mocked them with the accusation that they had been “bought with money." Even in the absence of explicit verbal abuse, participants often perceived hostility through facial expressions and body language. These nonverbal cues conveyed judgment and exclusion, reinforcing participants’ feelings of being unwelcome. P6, for example, described the discomfort she experienced upon arriving in Taiwan due to her accent, a concern echoed by P14, who noted that people who speak differently from local Taiwanese often receive “condescending looks.” Reflecting on her experiences, P6 remarked:

\begin{quote}
    “Over time, I’ve managed to cope with such reactions, but I remain profoundly concerned for my child, especially with the possibility of them being labeled as a ‘second-generation immigrant’ and the potential harm associated with such labeling.”
\end{quote}

P4 shared insights into how her identity as a new immigrant was sometimes belittled. And P17 emphasized the silent judgments she often faced from young people, “even if they might not voice it, their facial expressions say it all.” 

Regardless of the status they had held in their home countries, most participants told us that it was a struggle to live in a country that did not feel like home. As P2 put it, 

\begin{quote}
    “Despite being considered successful and knowledgeable in my home country, everything changed after I arrived in Taiwan. To many people here, I’m simply an outsider, someone unfamiliar with their ways. It’s not that we lack ability, but the differences in habits, environment, and especially language make it difficult. Mandarin isn’t our mother tongue, and sometimes the way we think or express ourselves differs. In those moments, we can sense that others are looking down on us.”
\end{quote}

Separately, the participants cited the challenge of re-establishing themselves in a new society. Although two (P1, P15) told us they felt comfortable living in Taiwan, many more said that—irrespective of their accomplishments and other individual characteristics—they were perceived principally as outsiders, and that this led to feelings of being undervalued or misunderstood.

\subsubsection{Concerns that their children were or would be treated as “less than”}
Beyond the discrimination they themselves faced, participants were also deeply concerned about their children being mistreated due to their identity as “second-generation immigrant children.” P2 shared that when participating in parent-child activities in Taiwan, she often felt apprehensive and vulnerable, fearing that her unfamiliarity with local parenting norms and broader cultural expectations might lead others to judge her child negatively. P10 recounted an incident of discrimination that was particularly striking, in which her half-Vietnamese son was told by another child to “go back to Vietnam.” In response, she tried to comfort her son by affirming their Vietnamese identity as a source of pride. However, the emotional pain of both mother and child remained evident. Similarly, P8, distressed by comparable experiences, actively sought advice from colleagues with similar backgrounds on how to help her child develop a strong and healthy sense of identity. P4 underscored the difficulty of ignoring or downplaying a child’s mixed Taiwanese-Indonesian heritage. She spoke of her growing anxiety that her child would eventually be ridiculed or excluded, given a prevailing societal perception that having a SEA mother marked someone not only as "different," but as "less than."

\subsection{Creating Engagement Opportunities with Children’s Learning}
\subsubsection{Imparting culture within domestic settings}

Despite their various challenges discussed above, our sample of SEA immigrant mothers shared numerous instances where, within the confines of their daily household routines, they made concerted efforts to educate their children about their cultural traditions and heritage. 

One instrumental resource that aided their children’s acquisition of their own native languages was knowledge of their home country. Almost all the participants made a conscious effort to converse in their native languages with their children at home, with the majority (P1, P3-P5, P7-P11, P14) claiming to have seamlessly integrated those languages into daily-life scenarios. For instance, P1 said she frequently used Vietnamese for routine interactions, such as simple greetings or counting. 

To strengthen their children’s connections with their maternal heritage cultures, most of the sampled mothers utilized objects linked to their home country. P2, for example, drew parallels between local Taiwanese games and those in  Cambodia. When her children played tug-of-war, she mentioned its importance in Cambodian culture and highlighted its status as a globally recognized game drawn from the tradition they shared. 

P5 used Vietnamese food-themed playing cards to turn gameplay into a learning experience about various Vietnamese dishes. P4, meanwhile, introduced her child to the traditional Indonesian strategy game of Congklak, not only to sharpen mathematical skills but to acquaint with Indonesian culture. 

Just over half the participants (P2, P4, P7-P9, P11-P12, P14) regarded the preparation of traditional dishes from their native countries as a gateway to their children’s immersive cultural learning. For example, during festivals, P7 prepared Vietnamese dishes to highlight the distinction between her home country’s culture and that of Taiwan, and thereby deepen her child’s appreciation of the former. In short, most of the participants embedded cultural education within their daily parent-child interactions: adeptly utilizing everyday objects, games, and verbal interactions to create an environment conducive to heritage-language learning and other forms of cultural acquisition and preservation. 

\subsubsection{Unique mother-child interactions while learning}
Some immigrant mothers persistently developed and engaged in unique forms of learning that were exclusive to them and their children. P2, for one, told us that although the mathematical methodologies she had learned in Cambodia diverged from those used in Taiwan, some elements of mathematics were universal. She fondly recounted time spent with her elder daughter while the latter was bathing: 

\begin{quote}
    "Bath time is when we enjoy ourselves the most. We treasure that secret space—just the two of us, sharing conversations… We write on the fogged-up mirror and come up with simple math problems together."
\end{quote}

There, the fogged-up bathroom mirror became an improvised blackboard. Her child wrote math problems on it, which P2 attempted to answer. When she eventually solved these problems, P2 felt a deep sense of accomplishment.

For P9, the most joyous moments of parent-child learning occurred when she and her child lay in bed at night playing a game called Idiom Solitaire on a mobile app. During this activity, she was often amazed by her child’s proficiency in Chinese, and saw it as an opportunity to hone her own knowledge of the language. But even more important, to her, was that the child immensely enjoyed playing this game, often staying up late and expressing reluctance to sleep due to excitement about it. 

Some mothers also described their efforts to better understand and support their children during times of academic stress. For instance, P13 joined a Facebook community where she connected with a Vietnamese educator who had studied parenting in the U.S. She applied what she learned from the online courses to daily communication with her child. P3, P4, and P12 shared that the time spent walking their children to and from school became valuable opportunities to connect. During these moments, they listened to their children share about what they had learned that day and offered words of encouragement.

Importantly, these types of unique connections typically arose in moments when the mother and child were alone together, without other people’s presence. This is perhaps unsurprising, in light of the above-noted resistance by SEA mothers’ Taiwanese family members.

\section{Discussion}
\textcolor{black}{
Our findings reveal that the marginalization of Southeast Asian (SEA) immigrant mothers in Taiwan is not merely a matter of cultural adjustment or digital literacy, but a symptom of a broader political economy of care. As scholars have noted, the rise of transnational marriage in East Asia is structurally linked to a "care deficit," where immigrant women are recruited to fill gaps in reproductive and caring labor \cite{liang_child-rearing_2020}. These mothers enter Taiwan through a system that legally defines them as spouses but socially positions them as utilitarian caregivers \cite{wang_being_2010}.}

\textcolor{black}{This structural positioning creates a "double bind": the state relies on them to raise "New Taiwanese" children, yet simultaneously marginalizes the foreignness that defines their identity. To address these harms, we propose the Intersectional Justice-Oriented Design framework. We position this framework primarily to prescribe actionable directions for HCI, grounding these prescriptions in our empirical description of SEA immigrant mothers' lived experiences and our diagnosis of structural conditions. Thus, the framework serves as a scaffold to support researchers and practitioners in designing technology for these complex environments.}

\subsection{Pursuing Justice through Intersectional Identities}
Our findings point toward deeper socio-technical barriers embedded in individual, domestic, and societal relationships. Applying Chordia et al.’s discussion around harm allows us to foreground how marginalization operates not only through absence of resources, but through relationships of silence, judgment, and self-erasure. It is also important to raise awareness to SEA immigrant mothers’ intersectional identities  \cite{kirabo2021inclusive, crenshaw2017intersectionality, Schlesinger2017}, as Chordia et al. suggested  \cite{chordia_social_2024}, to “reveal uniquely felt forms of harm” that may not be otherwise recognized in the Global North context. Rather than simply categorizing harm, we use this framework to interrogate how design can engage with such intersectional identities of SEA mothers’ everyday lives—textures often invisible in conventional HCI accounts of parent—child interactions. We interpret our findings through three forms of harm that are particularly salient in the lived experiences of SEA immigrant mothers in Taiwan: cultural harm, recognition harm, and psychological and emotional harm. These harms are often overlapping and mutually reinforcing, yet conceptually distinct.

To address the harms of marginalized groups, Chordia et al. propose three levels of intervention: individual, community, and structural \cite{chordia_social_2024}. This intervention framework aligns closely with our data analysis, which identifies the challenges SEA mothers encounter in engaging with their children’s learning activities across three levels: individual, familial, and societal. We also draw upon Dombrowski et al.’s six design strategies for social justice: designing for transformation, recognition, reciprocity, enablement, distribution, and accountability \cite{dombrowski_social_2016}. Of these, we \textcolor{black}{recognize} that three strategies—designing for recognition, reciprocity, and accountability—can be mapped onto and leveraged to support immigrant mothers across these levels \cite{dombrowski_social_2016}.

By bringing these two strands of scholarship into dialogue, we build a conceptual foundation for what we term \textit{Intersectional Justice-Oriented Design}. This framing aligns the scales of justice-oriented intervention with design strategies that respond to the lived realities of intersectional harm. At the individual level, designing for recognition can restore dignity and presence for those whose contributions are routinely dismissed. At the familial level, designing for reciprocity creates infrastructures for mutual validation and intergenerational exchange. At the systemic level, designing for accountability enables collective mechanisms to document, challenge, and redress institutional exclusion. Together, this alignment provides a transferable framework for justice-oriented HCI that is responsive to the layered, intersectional nature of marginalization and capable of guiding socio-technical systems toward structural transformation.

\subsubsection{At the Individual Level: Designing for \textcolor{black}{Epistemic and Competence} Recognition}

At the individual level \cite{chordia_social_2024}, \textcolor{black}{ SEA immigrant mothers experience a profound crisis of \textit{epistemic authority} \cite{Ajmani2025} —their status as "knowers" is systematically undermined. While Sohn et al. \cite{Sohn2012Examination} argue that household cooperation relies on the "perceived capability" of family members to execute tasks correctly. Our findings show that SEA mothers are often perceived as lacking this capability simply because their knowledge is situated in a different cultural framework. For instance, P2’s child thought she did not understand the school communication book , and P5 struggled to help with math because the "formulas are shown in a different way" . The harm here is is the erasure of competence.}

\textcolor{black}{Current social justice frameworks often focus on "giving voice" (recognition of identity). However, given the political economy of care where these mothers are valued primarily for labor, "voice" might not be sufficient. We must design for competence visibility. As Crabtree and Tolmie \cite{Crabtree2016Day} note, domestic order is maintained through the "methodical assemblage" of objects and routines. Design should help mothers insert their own "methodical assemblages" (their cooking, their math skills, their caring work) into the family's visible routine, converting their invisible background labor into recognized expertise.}

\textcolor{black}{Building on this, we argue that design must focus on epistemic and competency recognition—translating the mother’s invisible skills and knowledge into recognized assets. Design can support this by acting as a translator for competency. For example, rather than a tool that translates a math word problem (which keeps the mother passive), a system could explain the underlying concept using the educational system from the mother's home country. This allows the mother to teach the logic to her child, repositioning her from a confused bystander to an intellectual resource. By archiving and validating her competence, whether in math, cooking, language, and culture, technology can generate a digital footprint of expertise.}

\subsubsection{At the Familial Level: Designing for \textcolor{black}{Subversive} Reciprocity}

\textcolor{black}{“Subversive Reciprocity” is a critical concept that describes practices of mutual exchange that resist or disturb dominant power relations while still operating in collaborative ways. It is a critical concept that describes practices of mutual exchange that resist or disturb dominant power relations while still operating in relational ways.  At the familial level, the home functions not as a neutral domestic space but as a political site governed by patriarchal surveillance. Participants like P5 and P9 reported being silenced by in-laws when attempting to teach their native language or assign chores. In this context, current technology tools that assume a democratic household (e.g., shared calendars and to-do lists) may inadvertently expose mothers to conflict.}

\textcolor{black}{To navigate these constraints, we adopt Sultana et al.'s call for "Designing within the Patriarchy. \cite{sultana_design_2018}" Sultana et al. argue that in deeply patriarchal contexts, "activist" designs that aim to disrupt power structures directly can endanger women or result in the technology being banned by husbands and in-laws. Instead, they advocate for strategies that "empower within" and support situated tactics, accepting structural constraints temporarily to secure meaningful agency for women.}

\textcolor{black}{Applying this lens, we propose a direction that merges subversive reciprocity with Sultana’s call for safety and strategic alignment \cite{sultana_design_2018}. They suggest that technologies are more likely to be accepted if they package empowerment alongside content that husbands and in-laws already value. Since the political economy of care highly values the mother’s role in ensuring the child's academic success and discipline \cite{MotherhoodandEducation}. Here, a tool that attempts to address the language barrier and cultural transmission \cite{To2023, Liaqat2021Intersectional} may be rejected by the in-laws. We propose integrated scaffolding systems that strategically reframe the SEA immigrant mother’s cultural background and knowledge as indispensable assets for the child’s holistic development. Rather than viewing her heritage as a barrier, these systems position her educational history and cultural values as essential resources for ensuring the child's academic success and moral cultivation (e.g., leveraging her cultural emphasis on diligence and respect to reinforce behavioral goals valued by the household). By aligning with the Taiwanese family’s priority (academic achievement and child discipline), the design leverages the patriarchal value system to legitimize her cultural authority, making her participation "sanctioned" rather than overtly subversive in the familial context.}

\textcolor{black}{In addition, Sultana et al. emphasize shifting focus from "solving problems" to supporting the "situated tactics" women already use to cope, such as forming peer groups away from men. Our findings revealed the "fogged-up mirror" as a distinct situated tactic—a fleeting, private moment where the mother claimed space to teach her child. We argue for designing micro-territories of safety—digital equivalents of the fogged-up mirror. These are ephemeral and asynchronous channels (e.g., disappearing voice notes, private digital vaults) that allow mothers to transmit cultural heritage without leaving a permanent digital footprint visible to in-laws. This supports "Subversive Reciprocity" by prioritizing the safety of the mother-child bond over the transparency of the household, allowing mothers to assert agency within the patriarchal structure.}


\subsubsection{At the Societal Level: Designing for \textcolor{black}{Structual} Accountability}

On a societal level \cite{chordia_social_2024}, \textcolor{black}{we must address the "trust deficit" imposed on these mothers. Recent work by Hsiao et al. introduces "sociotechnical adaptation," describing how migrants adjust their technology use to build trust and eventually gain offline belonging \cite{Hsiao2023How}. The "organic" trust-building Hsiao et.al. describe is blocked by structural racism in SEA immigrant mothers. Our findings suggest this "adaptation" loop is fractured for SEA mothers. Because they are stigmatized as "commodities" (e.g., being mocked as "bought" ) and culturally inferior,  it is difficult to assume individual adaptation can overcome the structural bias of the host society. For instance, P2 and P4 reported being mocked in public as having been "bought," and P10’s child was told to "go back to Vietnam". In this relatively hostile context, the burden of building trust cannot be placed solely on the migrant to "adapt"; the sociotechnical system must instead demand accountability from the host society \cite{dombrowski_social_2016}.}

\textcolor{black}{While Hsiao et al. emphasize building a "digital footprint" to prove trustworthiness to locals, our participants face a different challenge: their "footprint" is actively erased or devalued. Their cultural knowledge is treated as inferior to the dominant Taiwanese curriculum. Therefore, \textit{Intersectional Justice-Oriented Design} at the societal level must shift from facilitating connection (which assumes equality) to establishing \textit{epistemic citizenship} \cite{chordia_social_2024, crenshaw2017intersectionality, FRASER2021113817}. This means designing systems that render the mothers' invisible labor (cultural teaching, caregiving) publicly visible and valuable, thereby challenging the narrative that they are merely passive beneficiaries of Taiwanese benevolence. Instead of building "connection" (which assumes equal footing), we should create mechanisms to report bias and document exclusion, forcing the "offline" society to be accountable for the harm that prevents "online" trust. The goal is not just to help mothers "fit in," but to force the host society to recognize their value. We propose designing for \textit{Structural Accountability} through mechanisms of collective witnessing.}

\textcolor{black}{This could take the form of aggregated, anonymous reporting platforms where mothers document instances of institutional bias (e.g., discrimination in schools or public spaces). Unlike individual complaints, which can be dismissed, visualizing this data as collective "heatmaps" of exclusion renders the "unfriendly social environment" visible and undeniable. Furthermore, platforms could support the creation of "living libraries" of Southeast Asian heritage, formally archiving the mothers' cultural knowledge as a public resource. By converting their private domestic labor into public cultural capital, these systems challenge the deficit model and force the wider society to encounter these women as educators and experts, not just wives.}

\subsubsection{ \textcolor{black}{Intersectional Justice-Oriented Design Framework}}

\textcolor{black}{Ultimately, these extensions illustrate that while the \textit{Intersectional Justice-Oriented Design} framework is grounded in the lived experiences of SEA immigrant mothers in Taiwan, it offers a transferable conceptual orientation for operationalizing intersectionality across diverse contexts of marginalization. By guiding design practices to address layered forms of oppression, from marriage migrants navigating Global South-to-North transitions to refugees and indigenous communities facing structural erasure, the framework demonstrates broad utility. In each scenario, its core tenets remain essential yet contextually adaptive: competency recognition affirms devalued knowledge and identity; subversive reciprocity addresses patriarchal surveillance within the domestic context; and structural accountability to establish epistemic citizenship. Thus, the framework serves not only as a lens for our specific case study but as a versatile, generative tool for HCI researchers and designers to advance justice-oriented design and build more equitable socio-technical systems beyond any single geographic setting.}

\begin{figure*}
  \centering
  \includegraphics[width= 0.85\linewidth]{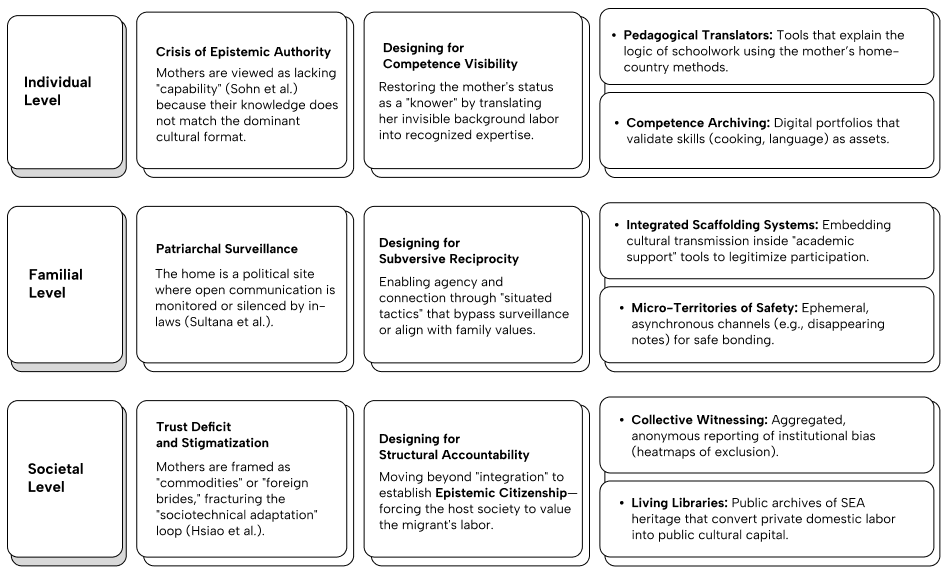}
  \caption{The Intersectional Justice-Oriented Design Framework}
  \Description{the Intersectional Justice-Oriented Design framework}
\end{figure*}

\subsection{Flourishing \textcolor{black}{ Is A Measure of Justice}}

\textcolor{black}{By converting SEA immigrant mothers' private domestic labor into public cultural capital, the systems mentioned above challenge the "deficit model" and force the wider society to encounter these women as educators and experts, not just mothers and wives.
However, justice is not solely the absence of harm or the rectification of structural wrongs. To stop at accountability would be to define these mothers only by the oppression they face. Therefore, we must extend our framework to encompass the ultimate goal of these interventions: the capacity to thrive on one's own terms.}

\textcolor{black}{In the context of the political economy of care, SEA mothers are often reduced to their utility as laborers. To counter this, we argue that justice-oriented design should prioritize flourishing \cite{To2023}, defined not merely as happiness, but as the restoration of agency, cultural pride, and future-oriented aspiration.}

\textcolor{black}{Building on To et al.’s \cite{To2023} critique of "damage-centered" design, we assert that addressing the harms of patriarchal surveillance and epistemic erasure is only the baseline. True justice requires systems that actively cultivate "joy as resistance. \cite{Bosley2022}" When a mother uses a "fogged-up mirror" to teach math, she is not just "surviving" a hostile home; she is engaging in a creative, generative act of parenting. Technology should amplify these moments, moving beyond "support tools" (which imply deficiency) to "celebratory infrastructures" \cite{Grimes, Fronemann} that treat the mother’s cultural heritage as a source of abundance rather than a barrier to be overcome.}

\textcolor{black}{This shifts the fundamental question of HCI design for immigrant families. Rather than asking, \textit{"How can we help these mothers catch up to the Taiwanese standard?"} as a deficit question, a flourishing-centered approach asks, \textit{"What do these mothers hope for? How can technology archive their pride, validate their lineage, and allow them to define success on their own terms?"} By centering flourishing, we position SEA immigrant mothers not as victims of a care deficit, but as architects of a new, pluralistic Taiwanese identity."}

~
\section{Limitations and Future Work}
This study offers important insights into the experiences of SEA immigrant mothers in Taiwan, but several limitations should be noted. First, our participants had at least basic Mandarin proficiency and had lived in Taiwan for over eight years. While they reflected on earlier language barriers and changing experiences over time, future research can explore the perspectives of recently arrived immigrant mothers, who may face different challenges. Second, the sample size was relatively small and predominantly included Vietnamese mothers. Although previous research (e.g., Ambe et al. \cite{Ambe2019}) highlights the value of in-depth data from small hard-to-reach groups, future studies could include more SEA immigrant mothers across countries and languages. Future work could include other family members, such as fathers, and adopt longitudinal approaches to better understand how families adapt over time in evolving socio-cultural contexts.

Finally, while this framework outlines specific design actions, we acknowledge that technology is not a panacea for systemic injustice. A design intervention cannot overturn immigration policies or deep-seated patriarchal norms. We caution against 'solutionism' that might treat these tools as fixes for structural violence. Instead, we view these designs as supportive infrastructures—mechanisms that provide necessary resources and validation for mothers to sustain their agency, even while broader structural challenges persist

\section{Conclusion}
\textcolor{black}{In examining parent-child interaction through the lived experiences of SEA immigrant mothers in Taiwan, our work highlights how everyday learning is shaped not merely by digital literacy or cultural difference, but by the political economy of care. We expose how the structural positioning of these mothers, recruited to fill a reproductive labor deficit yet stigmatized as "foreign brides," creates profound epistemic and relational harms that render their cultural authority invisible.}

\textcolor{black}{By surfacing these harms, we offer \textit{Intersectional Justice-Oriented Design} as a framework for structural responsiveness in HCI. We argue that to support marginalized caregivers effectively, design must move beyond "empathetic support" toward specific political interventions: At the individual level, we must design for competence visibility, creating tools that translate invisible domestic labor into recognized epistemic authority. At the familial level, we must enable subversive reciprocity, providing ephemeral and safe channels for mothers to assert agency within the cracks of patriarchal surveillance. At the societal level, we must design for structural accountability, establishing mechanisms for collective witnessing that challenge the "trust deficit" imposed by the host society.}

\textcolor{black}{Ultimately, this work contributes to immigrant studies and social justice in HCI by demonstrating that "inclusion" is an insufficient goal when the system itself is exclusionary. Instead, we advocate for a design agenda centered on flourishing, one that validates the "situated tactics," cultural aspirations, and resilience of immigrant mothers, positioning them not as passive recipients of aid, but as active architects of their own family narratives.}

\begin{acks}
We are deeply grateful to the Southeast Asian immigrant mothers who participated in this study. We thank them for welcoming us into their lives and for their courage in sharing their lived experiences of parenting and migration. We appreciate Dr. Sharifa Sultana for her valuable suggestions on revising this work. We also acknowledge the support of the National Science and Technology Council (NSTC), Taiwan.
\end{acks}

\bibliographystyle{ACM-Reference-Format}
\bibliography{citation}

\appendix
\section{Appendix: Interview Protocol}
\subsection{First Semi-structured Interview}

This first interview protocol was designed to explore how Southeast Asian immigrant mothers in Taiwan engage in their children’s learning at home, and what factors shape or constrain such involvement. The questions aim to surface mothers’ personal experiences, daily routines, learning interactions, aspirations, and challenges in domestic contexts. All interviews were conducted in Mandarin. The protocol below is translated from Chinese to English.

~
\subsubsection*{Warm-up and Background}
\begin{enumerate}
    \item [1. ] To begin, could you tell me about yourself and your family?
        \begin{itemize}
            \item Could you introduce the members of your household? 
            \item Could you share the story of how you came to live in Taiwan?
        \end{itemize}
    \item [2. ] Where are you from, and what was your educational background in your home country?
        \begin{itemize}
            \item How did you learn Mandarin and adjust to life in Taiwan?
            \item Could you share a memorable story from your experience adapting to Taiwanese society?
        \end{itemize}
\end{enumerate}

\subsubsection*{Daily Routines and Learning Activities}
\begin{enumerate}
    \item [1. ] Could you describe your child’s typical daily routine from morning to bedtime?
        \begin{itemize}
            \item For example, what did yesterday look like for your child?
            \item In your view, which parts of the day count as “learning” and which are more about rest or play? Why?
            \item How do you usually participate in these activities?
        \end{itemize}
    \item [2.] Can you tell me about your family life at home?
        \begin{itemize}
            \item What kinds of activities do you usually do together as a family?
            \item Which of these activities do you consider part of your child’s learning, either academic or life learning?
            \item Could you walk me through how you and your child do these activities together?
        \end{itemize}
    \item [3.] What does an ideal learning environment look like for your child at home?
        \begin{itemize}
            \item What are the ways you help your child achieve this ideal? How is technology involved in this process?
        \end{itemize}
    \item [4. ] How do you seek resources to support your child’s learning at home?
    \item [5. ] What do you find most difficult or challenging when engaging in learning activities with your children at home?
        \begin{itemize}
            \item Could you give a concrete example?
            \item What strategies do you use to address these challenges?
            \item What tools, information, or people do you rely on to navigate those moments?
        \end{itemize}
    \item [6. ] What is one of the most enjoyable learning experiences you've had with your child? 
    \item [7. ] Could you draw or list the objects or technologies you and your child use during learning?
        \begin{itemize}
            \item How do you use each one together?
            \item Have you experienced any difficulties in using these tools? How did you address them?
            \item Have any tools been helpful in supporting children’s learning? Are there still any needs or challenges that these tools have not been able to meet?
        \end{itemize}
\end{enumerate}

\subsection{Second Semi-structured Interview}
The second interview aimed to follow up on the diary entries and deepen our understanding of participants’ reflections on everyday learning activities with their children. The conversation focused on specific episodes described in the diaries, parenting aspirations, emotional moments, and household decision-making processes. All interviews were conducted in Mandarin. The protocol below is translated from Chinese to English.

\begin{enumerate}
    \item [1. ] In your diary entry on \_\_ [Month/Day], you wrote about \_\_ [a specific interaction or activity] with your child. Could you tell me more about that experience in detail?
    \item [2. ] Over the past two weeks, what was the most fulfilling moment for you in relation to your child’s learning?
        \begin{itemize}
            \item Could you share more details about what happened and how you felt?
            \item What about the most frustrating or difficult moment?
            \item What role does technology play in this moment?
        \end{itemize}
    \item [3. ] How do you plan or manage such a full daily life for yourself and your child?
        \begin{itemize}
            \item How do you arrange a time to be together?
            \item Who else is involved in making these plans or preparing activities?
            \item Who usually makes the final decisions?
            \item How do these decisions affect your relationship with your child?
        \end{itemize}
    \item [4. ] In what ways do you share your home country’s language and culture with your child?
        \begin{itemize}
            \item Are there particular activities, stories, or practices that you like to do together?
        \end{itemize}
    \item [5. ] As a Southeast Asian immigrant mother in Taiwan, is there anything you’d like to share in this role?
    \item [6. ] How does it shape the way you interact with your child?
\end{enumerate}

\section{Appendix: Codebook}
~
{\color{black}
{\small
\onecolumn\begin{longtable}{>{\RaggedRight\arraybackslash}p{3cm} >{\RaggedRight\arraybackslash}p{3.6cm} >{\RaggedRight\arraybackslash}p{6.8cm}}
\caption{Overview of the Codebook}
\label{tab:codebook} \\
\toprule
\textbf{Category} & \textbf{Codename} & \textbf{Definition} \\
\midrule
\endfirsthead

\toprule
\textbf{Category} & \textbf{Codename} & \textbf{Definition} \\
\midrule
\endhead

\midrule
\multicolumn{3}{r}{{Continued on next page}} \\
\midrule
\endfoot

\bottomrule
\endlastfoot

Learning Activities and Engagement at Home & Accompanying Children in Homework & Mothers stay nearby while children do homework, mainly for supervision. \\
& Supporting Language Learning with Technology & Mothers help children learn Mandarin, English, or their native language with digital tools. \\
& Doing Household Chores & Mothers assign and share chores with their children, as chances to help building practical life skills. \\
& Playing and Extracurricular Activities & Mothers support children's participation in games, outdoor play, and physical activities. \\
& Teaching Values in Daily Life & Mothers use everyday situations to help children practice values like honesty and kindness. \\

Challenges Faced in Supporting Children’s Learning & Struggling with Educational Support Due to Language Gaps & Mothers have trouble understanding school materials, leading to lower confidence and reliance on others for help. \\
& Feeling Dismissed in Parental Roles & Mothers feel excluded or undervalued in supporting their children’s learning, as children often turn to fathers or native speakers instead. \\
& Experiencing Opposition from Family Members & Mothers encounter resistance from husbands or in-laws who limit their parenting autonomy, including discouraging the use of native languages. \\
& Facing Societal Bias and Cultural Stereotypes & Mothers experience social exclusion or stigma tied to their immigrant identity, often feeling devalued compared to how they were perceived in their home countries. \\
& Worrying About Children’s Social Belonging & Mothers express concern that their children may face exclusion or identity struggles at school due to their mixed heritage or immigrant background. \\

Engagement Strategies and Cultural Agency & Creating Cultural Learning Time at Home & Mothers intentionally integrate cultural education into daily life through language, food, games, and conversations to preserve connections to their heritage. \\
& Developing Personalized Learning Moments & Describe creative, intimate mother-child learning experiences in private settings, fostering emotional connection and mutual learning outside formal contexts. \\
& Seeking Information through Digital and Social Networks & Mothers access educational resources via online platforms and personal networks, like websites or online groups. \\

\end{longtable}
}}


\end{document}